\documentclass[useAMS,usedcolumn,eps,psfig,graphics]{mn2e}
\usepackage{epsfig, graphicx,natbib2}

\def\grtsim{\mathrel{\hbox{\rlap{\hbox{\lower2pt\hbox{$\sim$}}}\raise2pt\hbox{$>$}}}}
\def\lesssim{\mathrel{\hbox{\rlap{\hbox{\lower2pt\hbox{$\sim$}}}\raise2pt\hbox{$<$}}}}

\def\degree{\nobreak\ifmmode{^\circ}\else{$^\circ$}\fi}

\newcommand{\aap}{A\&A}

\newcommand{\aj}{AJ}
\newcommand{\apj}{ApJ}
\newcommand{\apjl}{ApJL}
\newcommand{\apjs}{ApJS}

\newcommand{\mnras}{MNRAS}
\newcommand{\nat}{Nat}

\begin{document}

\topmargin -0.5in 

\title[The simlar stellar populations of quiescent spiral and elliptical galaxies]{The similar stellar populations of quiescent spiral and elliptical galaxies}
\author[Aday R.~Robaina et al.]{Aday R.~Robaina$^{1}$\thanks{E-mail:arobaina@icc.ub.edu (ARR)}, Ben Hoyle$^{1}$, Anna Gallazzi$^{2}$, Raul Jim\'enez$^{1}$, Arjen van der Wel$^{3}$\newauthor{Licia Verde$^{1}$}\\
\footnotesize\\
$^{1}$Institut de Ciencies del Cosmos, ICC-UB, IEEC, Marti i Franques 1, 08028, Barcelona, Spain, \\
$^{2}$Dark Cosmology Centre, Niels Bohr Institute, Juliane Maries Vej 30, 2100 Copenhagen, Denmark\\
$^{3}$Max Planck Institute for Astronomy, K\"onigstuhl 17, D-69117, Heidelberg, Germany
}

\pagerange{\pageref{firstpage}--\pageref{lastpage}} \pubyear{}

\maketitle

\label{firstpage}

\pagerange{\pageref{firstpage}--\pageref{lastpage}} \pubyear{}

\maketitle

\label{firstpage}

\begin{abstract}

 We compare the stellar population properties in the central regions of visually classified non-starforming spiral and elliptical galaxies from Galaxy Zoo and SDSS DR7. The galaxies lie in the redshift range $0.04<z<0.1$ and have stellar masses larger than $logM_*=10.4$. We select only face-on spiral galaxies in order to avoid contamination by light from the disk in the SDSS fiber and enabling the robust visual identification of spiral structure. Overall, we find that galaxies with larger central stellar velocity dispersions, regardless of morphological type, have older ages, higher metallicities, and an increased overabundance of alpha-elements. Age and alpha-enhancement, at fixed velocity dispersion, do not depend on morphological type. The only parameter that, at a given velocity dispersion, correlates with morphological type is metallicity, where the metallicity of the bulges of spiral galaxies is 0.07 dex higher than that of the ellipticals. However, for galaxies with a given total stellar mass, this dependence on morphology disappears. Under the assumption that, for our sample, the velocity dispersion traces the mass of the bulge alone, as opposed to the total mass (bulge+disk) of the galaxy, our results imply that the formation epoch of galaxy and the duration of its star-forming period are linked to the mass of the bulge. The extent to which metals are retained within the galaxy, and not removed as a result of outflows, is determined by the total mass of the galaxy.


\end{abstract}


\section{Introduction}

Galaxies in the local Universe come, broadly speaking, in two flavors:  objects with blue and red optical colors tend to inhabit different regions of the color-magnitude diagram \citep[CMD,][]{strateva}, with blue galaxies showing a large spread in color and red galaxies following a relatively tight sequence. This so-called red sequence has been observed up to $z\simeq2$ and has grown in mass by a factor of $\sim 2$ since $z=1$, although the evolution in the massive end of the distribution remains controversial \citep{heav, bell04, cimatti, faber07, robaina10}. As galaxies with red stellar populations typically show low levels of star formation\footnote{A fraction of galaxies in the red sequence at all redshifts is red because of dust-obscuration rather than truly red and dead populations. This fraction decreases towards both lower redshift and high stellar mass.} (SF), the mechanism needed to add stellar mass to the red sequence must imply the migration of a certain number of objects from the blue cloud to the red sequence \citep{brinch, bell07, walcher} by quenching of their star formation.

 While galaxies on the blue cloud show predominantly disk-like light profiles, the red sequence is dominated by objects with more concentrated light distributions \citep{blanton03}. Further evidence on the relation between SF quenching and galaxy structure is provided by \citet{bell08}, who found that red and dead stellar populations tend to inhabit galaxies with concentrated light profiles. Detailed studies of the shape of quiescent galaxies show that spheroidal systems are overwhelmingly dominant at masses larger than $1-2 \times 10^{11}M_\odot$, but a large contribution of red disks is observed below that critical mass \citep{arjen09, masters}, in good agreement with the results by \citet{bundy10} on the migration of disk galaxies to the red sequence. More recently, \citet{holden11} report that at all redshifts $z<1$ the relative fraction of disk and early-type galaxies added to the red sequence at a given stellar mass is approximately constant. Taking all these previous results together, it is clear that SF quenching in disk galaxies without the need of dramatic morphological perturbations is a valid and frequent mechanism --although not dominant-- to move galaxies from the blue cloud to the red sequence.

For this reason, red disk galaxies have drawn the attention of the extragalactic community. In the late 70's, \citet{vdb} reported the existence of passive galaxies with spiral morphology in the Virgo Cluster, and later studies confirmed the existence of a population of quiescent disk galaxies in dense environments \citep[i.e.,][]{poggianti}. More recently \citet{wolf09} show that in intermediate-mass cluster environment red spiral galaxies are equivalent to the actively star-forming blue spirals, but with lower SF and a higher fraction of dust-obscuration. These galaxies also tend to display stronger bar features than their blue counterparts \citep{ben11a}. A hint on the origin of these systems is provided also by \citet{bamford} and \citet{skibba09}, who found by using visually classified SDSS galaxies from Galaxy Zoo (GZ) that the relation between optical color and environment is more significant than the well known morphology-density relation \citet{dressler}.

Decades of studies have shaped a solid knowledge of the stellar populations of 
red-sequence galaxies. These galaxies follow 
several tight scaling relations linking their stellar population properties  
to their mass and their dynamical and structural properties, such as the color-magnitude 
relation  \citep{faber73,GFW93}, the relation between absorption index strengths and 
velocity dispersion \citep{Bender93,Kelson06,Chang06} and the Fundamental Plane 
\citep{DD-FP,Bernardi03}.
The physical drive of these relations is an increase in all of metallicty, 
element abundance ratios and stellar age with galaxy mass or velocity dispersion 
\citep{Trager00,kunt01,thomas05,anna06, tojeiro10}. Indeed, stellar 
population properties seem to be more fundamentally correlated with stellar velocity 
dispersion than with galaxy mass \citep{Graves09}. The picture that emerges is that 
present-day elliptical galaxies with deeper potential 
wells have reached a higher degree of chemical enrichment and have formed their 
stars at earlier epochs and on shorter timescales. Moreover, the small intrinsic  
scatter in the observed scaling relations is associated primarily with variations  
in stellar age and, to a lesser degree, in chemical abundances, putting additional 
constraints on the variety of SFHs that present-day elliptical galaxies of similar 
mass have undergone \citep[e.g.][]{raul05,raul07,anna06,Graves10}. 

An additional parameter that influences the SFH, hence the stellar populations, of 
elliptical galaxies is their environment. While the slope of the scaling relations 
is independent of the environment, small variations in their zero-point and scatter 
have been observed, indicating both that the fraction of galaxies with younger 
stellar populations (``rejuvenated'') increases in low density environments 
\citep{Thomas10} and that at fixed mass galaxies in denser environments tend to be 
older than their low-density couterparts \citep{Cooper10,Clemens06, ben11}. 

On the other hand relatively few works (see e.g. Proctor\&Sansom 2002, Thomas\&Davies 2006, 
Kuntschner et al 2010, Falcon-Barroso et al 2011) have analysed the stellar populations and 
scaling relations of different morphological types, in particular among red-sequence galaxies.
Thomas\&Davies 2006 reanalysed the sample of spiral bulges (from Sa to Sbc) of Proctor\&Sansom 2002
 and found that the bulges of spiral galaxies 
have similar stellar populations to elliptical galaxies at fixed velocity dispersion. 
Early-type spiral galaxies also seem to follow the same Fundamental Plane 
as ellipticals, albeit with larger scatter (Falcon-Barroso et al 2011).

However these works generally do not distinguish galaxies on the basis of their star formation 
activity. In this work we are specifically interested in comparing the stellar populations 
of galaxies that are quiescent but differ in morphology, namely quiescent spirals against 
elliptical galaxies. \citet{masters} primary focus was in the characterization of red spirals and the comparison with blue spirals, but a detailed comparison of the stellar populations in quiescent spiral and early-type galaxies could shed some light on the processes by which they are formed and subsequently quenched. In particular, given the differences found between red and blue spirals in \citet{masters}, it would be extremely important to learn whether the stars in spiral galaxies can follow an evolutionary path similar to those in spheroidal systems even when the morphological evolution of their host galaxies is dramatically different, as that would put constraints on the mechanisms driving the star formation histories of passive galaxies in the Universe.

In this paper, we study the stellar populations in the central regions of a sample of truly passive spiral galaxies at $z\lesssim 0.1$ from SDSS and compare them to those in quiescent ellipticals. We choose to do so by comparing the ages, total metallicities ([Z/H]) and, in particular, the $\alpha-$enhancement of their populations. In order to assemble a statistically significant galaxy sample we use data products from the NYU-VAC \citep{blanton03} and visual morphology estimates from the Galaxy Zoo project \citep{lintott1,lintott2}. We also model the stellar populations in SDSS DR7 quiescent galaxies, following the method described in Gallazzi et al.~(2005, 2006)(hereafter G05 and G06 respectively), to obtain stellar masses, r-band weighted ages, [Z/H] and $\Delta$Mgb/$\langle$Fe$\rangle$ --a tracer of the $\alpha$-enhancement. We end up with a sample of $\sim$1000 quiescent spiral and $\sim$14700 passive early-type galaxies.

This paper is organized as follows: In Section 2 we describe the sample selection and the parameters we use to characterize the stellar populations. In Section 3 we present our results and discuss possible evolutionary paths. Finally, in Section 4, we present our conclusions.

Throughout this paper we use $\Omega_{m0}$=0.3, $\Omega_{\Lambda 0}$=0.7 and $h_{100}$=0.7. All magnitudes are in the AB photometric system.

\section{Sample selection and method}

Galaxies are drawn from the Sloan Digital Sky Survey Data Release 7 \citealp[SDSS DR7,][]{dr7}. In particular, we make use of the publicly available New York University-Value Added Catalog (NYU-VAC) released by \citet{blanton05}. We select galaxies in the redshift range $0.04<z<0.1$ and masses log $M_*>10.4 M_\odot$. The magnitude limit in the SDSS spectroscopy places the lower mass limit for red galaxies at $z\sim 0.1$ approximately at log$M_*=10.6 M_\odot$, although we choose to work with galaxies slightly below that limit. Nonetheless, we warn the reader that our sample is incomplete by $\sim$20\% below log $M_*=10.6 M_\odot$ at the highest redshift probed here. We make use of NYU-VAC k-corrected photometry, spectroscopic redshifts and light-profile fitting parameters in the r band.

Stellar masses, metallicities and r-band light-weighted ages for SDSS DR7 galaxies have been estimated in the same way as for previous releases and as described in G05, to which we refer the reader for a full description. Briefly, estimates of the stellar population parameters are obtained by comparing the observed stellar absorption features (corrected for emission lines) with those predicted by a comprehensive library of model spectra based on \citet{bc03} SSPs convolved with Monte Carlo star formation histories. A comparison between the new SDSS DR7 parameters and those of SDSS DR4 from G05 and G06 provides no systematic offset and a typical dispersion at the level of $\sim 0.1$ dex in light-weighted age, stellar mass and metallicity for galaxies with red, {\it quiescent} stellar populations, well below the typical error budget in those measurements. In this work we correct the estimated galaxy ages to z=0 by adding the lookback time at the redshift of the galaxy under the assumption of passive evolution (which is very reasonable for our sample of quiescent galaxies).

In addition to the aforementioned stellar population parameters we focus on the $\alpha-$enhancement. In spite of the name, the effect is more a lack of iron rather than an excess of $\alpha-$elements. This lack of iron is produced when the SF timescale of a galaxy is short. Core-collapse supernovae enrich the medium with $\alpha$-elements in scales of a few tenths of Myrs, while the Fe-enrichment is due to type Ia supernovae explosions. If a significant fraction of the stars are formed in a period shorter than the $\sim 1$ Gyr needed by type Ia supernovae progenitors to evolve, the stars would show a chemical composition with higher $\alpha$/Fe abundance ratios than those in stellar populations formed with longer timescales. 

As a tracer of the $\alpha$-enhancement in the stellar populations of our sample we use the semi-empirical [alpha/Fe] indicator adopted by G06, namely $\Delta$(Mgb/$\langle$Fe$\rangle$) which is the difference between the observed Mgb/$\langle$Fe$\rangle$ absorption index\footnote{$\langle$ Fe $\rangle$ is the average of the Fe5270 and Fe5335 index strengths} and that of the scaled-solar BC03 model that best fits [$\alpha$/Fe]-independent features. G06 have tested, through comparison with Thomas et al.~(2003) models with variable abundance ratios, that $\Delta$(Mgb/$\langle$Fe$\rangle$)) correlates linearly with the abundance ratio [$\alpha$/Fe] independently of age and metallicity (except for metallicities below 30 percent solar, which is lower than the range covered by our sample). In particular, we confirm
such proportionality over the metallicity and age range spanned by our sample
($0.5<Z/Z_\odot<2$ and age older than 3~Gyr) with the differential models presented
in \citet{walcher09} based on the theoretical \citet{coelho07} models
calibrated onto either BC03 or \citet{vazdekis10}. We stress  that the values  of  $\Delta$(Mgb/$\langle$Fe$\rangle$) should not be directly translated into values of [$\alpha$/Fe], i.e. the proportionality constant between  $\Delta$(Mgb/$\langle$Fe$\rangle$) and [$\alpha$/Fe] is not 1 and is likely model dependent.

\subsection{Quiescent stellar populations}

\begin{figure}
\begin{center}
\psfig{file=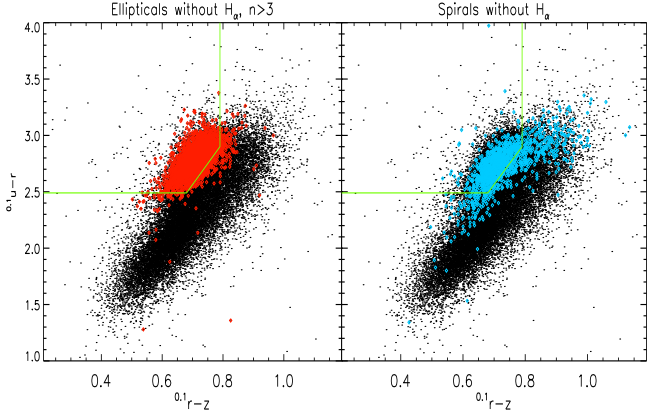,width=9cm,height=6cm,angle=0} 
\caption{Color-color diagram of SDSS DR7 galaxies with $log(M_*/M_\odot) > 10.4$ at $0.04<z<0.1$. {\it Left panel:} Elliptical galaxies with no $H_\alpha$ emission and sersic index $n>3$ are shown with red symbols. {\it Right panel:} Spiral galaxies with no $H_\alpha$ emission are shown with blue symbols. The green solid line in both panels defines the box used to select galaxies with quiescent stellar populations. This work concentrates on the comparison between objects shown as blue and red points within the selection box.
}
\label{fig:co-co}
\end{center}
\end{figure}

As we want to compare spiral and elliptical galaxies with {\it quiescent} stellar populations,-- and specifically not galaxies reddened by dust obscuration--, we select galaxies without $H_\alpha$ emission from their spectra. Nonetheless, two effects can endanger the reliability of this selection: a) large levels of dust obscuration might cause the absence of $H_\alpha$ detections even in the presence of star formation, and b) the $3\arcsec$ diameter of the SDSS fiber might not be large enough, in many cases, to sample a representative fraction of the stellar populations present in the galaxy. This would be specially concerning in the case of galaxies with quiescent bulges and star forming disks. These problems might be alliviated by the inclusion of an additional criterion. It has been shown that galaxies dominated by passive (i.e., non-star-forming) populations tend to be found in a particular region of a color-color plot in which one of the colors bracket the 4000$\AA$ break and the other falls redwards of that spectral feature \citep{williams, holden11}\footnote{A similar method but using a combination of optical and UV colors has been succesfully applied to the selection of red, old, disk galaxies by \citet{wolf09}}. In such a diagram red-passive and red-obscured galaxies are distinguished by the different imprint imposed by age and dust on the galaxy's  spectral energy distributions (SEDs). We show the SDSS derived $u-r$ color vs $r-z$ color diagram of the galaxy sample in Fig.~\ref{fig:co-co}. From the comparison between both panels, it is clear that elliptical and spiral galaxies without $H_\alpha$ emission present a different distribution. More than 90\% of elliptical galaxies lie on the region of the diagram where quiescent stellar populations cluster while spirals present a larger dispersion, both towards bluer $u-r$ color, probably indicating different stellar populations in the centre and outskirts of the galaxy, and towards redder $r-z$ values, implying larger levels of obscuration by dust. Therefore, we decide to use for the present study only those galaxies without noticiable $H_\alpha$ emission {\it and} quiescent stellar populations over the whole galaxy by selecting them in the $u-r$ vs $r-z$ diagram. The green line in both panels of Fig.~\ref{fig:co-co} shows the selection box used here. We have been very conservative in the definition of the boundaries between star-forming and non-star-forming. If we were to select galaxies at $\sim 1\sigma$ from the center of the quiescent clump, we would include several massive spiral galaxies with residual levels of star formation, detected at different colors in the disk and bulge in a visual inspection. The reason is that the galaxies in our sample span more than 1 dex in stellar mass, and the boundary between blue cloud and red sequence is mass- (or magnitude-) dependent. Then, we decide to include galaxies 0.04 mag redder in $u-r$ than the $1 \sigma$ boundary. This shift is arbitrary, but help us to make sure that the galaxies we study are, indeed, quiescent.

\begin{figure}
\begin{center}
\psfig{file=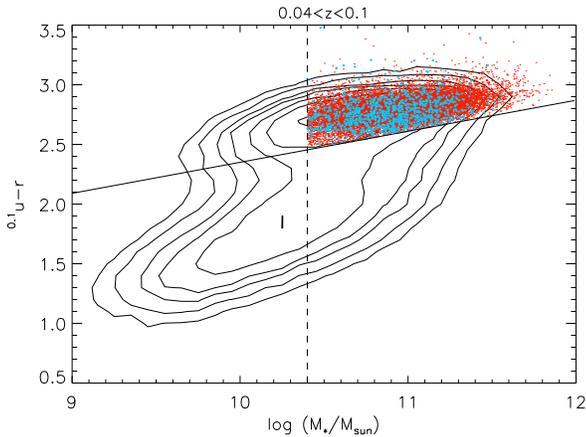,width=8cm} 
\caption{Color-stellar mass relation for SDSS DR7 galaxies at $0.04<z<0.1$. Contours show the density of galaxies in that redshift slice. Blue points represent our final sample of $\sim 1000$ quiescent spiral galaxies, while red points show our final sample of $\sim 19000$ quiescent ellipticals. Vertical dashed line corresponds to our adopted lower mass limit of log($M_*/M_\odot)=10.4$. Solid line corresponds to our adopted definition of red sequence galaxies.}
\label{fig:co-mass}
\end{center}
\end{figure}

Our sample selection differs from that in previous studies making use of Galaxy Zoo data \citep{masters} in the sense that we do not only select by optical color, but specifically reject any galaxy with hints of recent star formation.

\subsection{Systematic error checks}

\begin{figure}
\begin{center}
\psfig{file=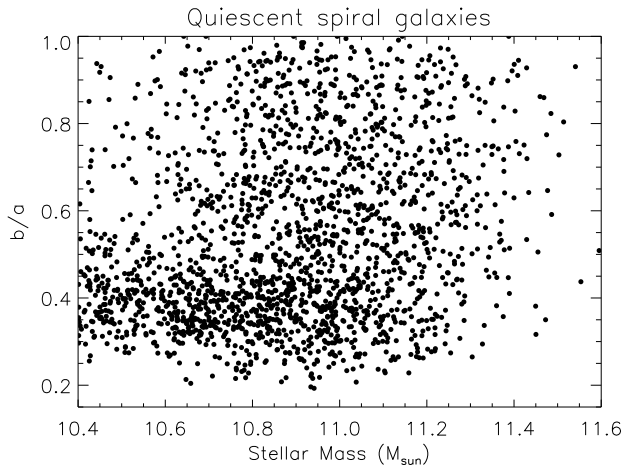,width=9cm} 
\caption{Axis ratio (r-band) vs galaxy stellar mass for our quiescent spiral galaxies as classified in Galaxy Zoo ($P_{CS}>0.8$). The majority of relatively low inclination ($b/a>0.5$), lower mass quiescent spirals are not confidently assigned to the ``spiral'' category in the Galaxy Zoo catalogue.}
\label{fig:spiral_q}
\end{center}
\end{figure}

The main aim of this work is to perform a differential analysis between the stellar populations of quiescent spiral and ellitpical galaxies. In order to identify these particular morphological types we use visual classifications released by the Galaxy Zoo (GZ) collaboration \citep{lintott1, lintott2}. Initally, we select galaxies with a debiased probability $P_{debiased}>0.8$ \citep{bamford} of being either spiral or elliptical (spiral arms or the presence of and edge-on disk is required in the GZ classification in order to assign a galaxy to the category of spiral). However, there are some relevant caveats one should take into account when using this catalog: a) quiescent galaxies have red optical colors and a higher M/L ratio than star forming galaxies; implying that for galaxies at the same mass, it is intrinsically more difficult to classify a red galaxy than a blue one, b) quiescent spiral galaxies usually lack the strong structure present in star forming spirals. This makes it even harder to recognize spiral patterns\footnote{Both \citet{masters} and the present study use galaxies with visible spiral patterns}, and c) the definition of ``Combined Spiral'' used in GZ catalog includes both face-on disk galaxies with obvious spiral arms {\it and} edge-on disks for which the spiral structure is not detectable even when present. The higher the redshift and the lower the stellar mass, the harder it is to clearly identify the spiral structure in a disk galaxy, even when present. All those problems could lead to a relatively high number of potential missclassifications, which are partially corrected for when calculating the debiased probability \citep[see][for further details]{bamford}. Nonetheless, as we show in Fig.~\ref{fig:spiral_q}, there is a clear bias present when studying quiescent galaxies. For a complete sample of spirals the projected axis ratio distribution should be approximately flat at values larger than the intrinsic thickness. We check for this by plotting the axis ratio (b/a) vs. galaxy stellar mass of the GZ-selected spirals with quiescent stellar populations. Truly face-on disk galaxies would fall close $b/a=1$, while those seen edge-on would appear somewhere in the range $0.1 < b/a < 0.5$ (depending on the intrinsic thickness of the disk and bulge contribution). Alternatively, in the absence of any bias the distribution with $b/a$ should be homogeneus at all masses. Instead we observe a 
deficit of low-inclination spirals at low masses. In order to guarantee the homogeneity of the spiral galaxy sample, we use only low-inclination galaxies ($b/a\ge0.5$). While this does not guarantee that we are complete down to our mass limit of $logM_*=10.4$, we can be sure that the morphological selection is homogeneus by including in the sample only those spiral galaxies with visible arms.

In the present work, we follow the Galaxy Zoo nomenclature, this is, we call `ellipticals' those galaxies which have a probability of being elliptical $>$0.8. However, this category includes many galaxies with B/T ratios lower than those of purely bulge-dominated objects and potentially, some misclassified face-on, smooth disks.

Large, massive, low-inclination spiral galaxies would be more likely to get an accurate classification than small, low mass counterparts. This is a matter of concern because potential systematic differences in the sizes and spiral structure of low- and high-inclination disk galaxies could jeopardize a proper comparison: the 3$\arcsec$ diameter of SDSS spectrograph fiber would sample systematically different physical radii in galaxies observed under different angles.

\begin{figure}
\begin{center}
\psfig{file=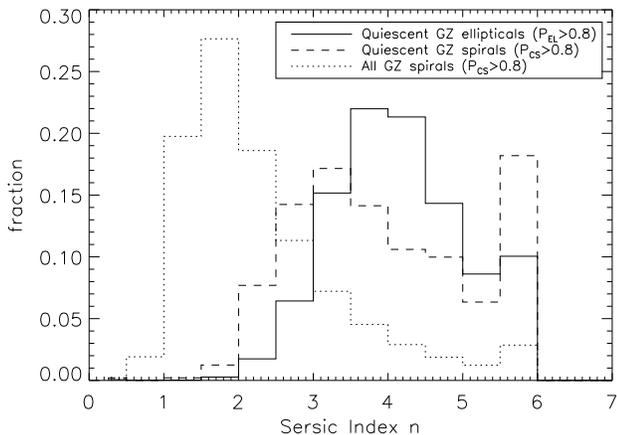,width=9cm} 
\caption{Sersic index distribution of quiescent elliptical, quiescent spiral and all spiral galaxies in the DR7 spectroscopic catalog. Only galaxies with debiased probability $P_{debiased}>0.8$ are selected in all three cases. Quiescent spiral galaxies show a distribution closer to that of early types than that of `normal' spirals.
}
\label{fig:sersic}
\end{center}
\end{figure}

We show the Sersic index distributions of quiescent ellipticals, quiescent spirals and all the spirals in our catalog, selected by a debiased probability $P_{debiased}>0.8$ of having the respective morphological type in Fig.~\ref{fig:sersic}. Notably, quiescent spiral galaxies present a different Sersic distribution than star-forming spiral galaxies --dominated by blue disks--, peaking at $n\simeq 4$ instead of lower value of $n\sim 1.5-2$ of blue spirals. This is indicative of the presence of large bulges in these objects. We note that introducing a further cut in $n$ to select elliptical galaxies (i.~e., selecting only elliptical galaxies with $n>3$) makes no difference to our final results.


Ideally, we would compare the stellar populations in the {\it bulges} of spiral and elliptical galaxies, as in the works by \citet{proctor02, thomas06}, but given the degeneracy between bulge-to-total ratio (B/T), physical size and angular distance evolution and the fixed size of the SDSS spectrograph fibre we will be probing the stellar populations in the {\it central regions}, that are indeed dominated by the bulge. Given the typical high sersic index of quiescent spirals, it is very likely that the light within $R_e$ is dominated by a prominent bulge. Nonetheless, we can not discard some contribution from stellar populations in the inner disk, although we deem it to be a second-order effect since we specifically select galaxies with quiescent stellar populations {\it all over} the galaxy. We will also test our main results for a subsample of galaxies for which we know the light in the fiber to be bulge-dominated.

While we focus in the comparison between elliptical and face-on spiral galaxies, we will also show in some of our plots a third group of galaxies, composed of edge-on quiescent disks selected from the NYU-VAC to have $b/a<0.4$ --and no constraint in GZ visual classification. When comparing this with the two aforementioned samples (edge- and face-on quiescent disks), the reader should bear in mind that in many cases the amount of light originated by the disks' stellar populations in edge-on objects will be larger than in the case of the GZ face-on quiescent spiral sample. 

As a summary of our sample selection, we use SDSS DR7 galaxies at $0.04<z<0.1$ with $log M_*/M_\odot > 10.4$. Objects are selected to show optical $u-r$ color compatible with red sequence galaxies, no $H_\alpha$ emission in the spectra. Additionally we select them photometrically to be quiescent in the $u-r$ vs $r-z$ diagram following \citet{williams} and \citet{holden11}. All objects must have debiased probability $P > 0.8$ of being either spiral or elliptical in the Galaxy Zoo catalog. Furthermore, we include only those spiral galaxies with low ellipticity ($b/a>0.5$). We will show for comparison purposes edge-on quiescent disk galaxies selected from the NYU-VAC to have $b/a <0.4$. This leaves us with a sample of approximately 14700 early type galaxies and 1000 face-on spirals.

\section{Results}

In this paper, we perform a {\it differential} analysis of the stellar populations in the central regions of quiescent spiral and elliptical galaxies. We compare, in particular, the metallicity, $\alpha-$enhancement (traced by the excess of Mgb/$\langle$Fe$\rangle$) and r-band light-weighted age of those two groups of galaxies over the redshift range $0.04<z<0.1$ and with stellar masses log$(M_*/M_\odot)>10.4$.

\begin{figure}
\begin{center}
\psfig{file=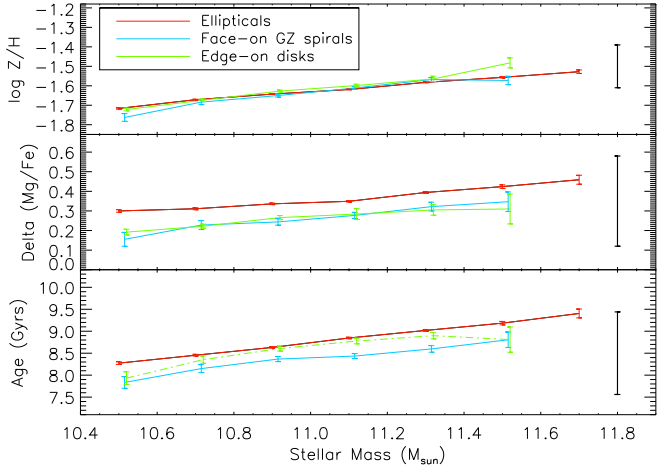,width=9cm} 
\caption{{\it Top panel:} Metallicity vs stellar mass. The metallicity in the central regions of quiescent spiral (dashed line) and elliptical (solid line) galaxies is identical. {\it Middle panel:} Excess of Mg$_b$ over Fe with respect to BC03 models. The $\alpha$-enhancement is very similar in the two populations. There is a small shift in the $\Delta$(Mg$_b$/$\langle$Fe$\rangle$) present in quiescent spirals with respect to ellipticals. The two groups are largely overlapping, as shown by the 1$\sigma$ error bar on the right, which represents the dispersion in the distribution. {\it Bottom panel:} r-band light-weighted age vs stellar mass. Again, quiescent spiral and elliptical galaxies are very similar, with a very weak trend of spirals showing slightly younger stellar populations. The large error bar in all three panels shows the typical 1$\sigma$ dispersion.}
\label{fig:vsmass}
\end{center}
\end{figure}

In Fig.~\ref{fig:vsmass} we show the median values of those quantities in bins of stellar mass. The metallicity in all three sub-samples is remarkably similar over the whole mass range explored in this paper. They are not only compatible within the $1 \sigma$ typical dispersion, shown as the large error bar on the right side of the plot, but also the position of the mean of both distributions are indistinguishable. The error in the position of the mean is denoted by the error in every mass bin.

While the fact that massive elliptical galaxies are $\alpha-$enhanced is well stablished in the literature \citep{wor92,thomas03,thomas05,anna06}, we find at very high statistical significance, that {\it at all masses}, the central regions of quiescent spiral galaxies also show an important excess of $\alpha-$elements with respect to Fe, and follow a trend with mass similar to that of the ellipticals. Both populations are largely overlapping, but in this case, the position of the mean seems to be shifted down by $\sim 0.1$. Light-weighted ages of quiescent spirals seem to be, on average, 400 Myrs younger than those of ellipticals. Nonetheless, errors in the measurements, modelling and derivation of parameters could add-up an error similar to such a small difference. If we are conservative and assume that the typical observational error in Mgb/$\langle$Fe$\rangle$ always work in the direction of increasing the recovered enhancement (i.e., if we consider only as true enhanced those galaxies with $\Delta$(Mgb/$\langle$Fe$\rangle$)$>$0.2), we can make sure that the bulk of elliptical galaxies at all the masses probed here possesses, indeed, a super solar $\alpha$/Fe ratio. Quiescent spiral galaxies above $10^{10.6}M_\odot$ are also inequivocally $\alpha$-enhanced.

\begin{figure}
\begin{center}
\psfig{file=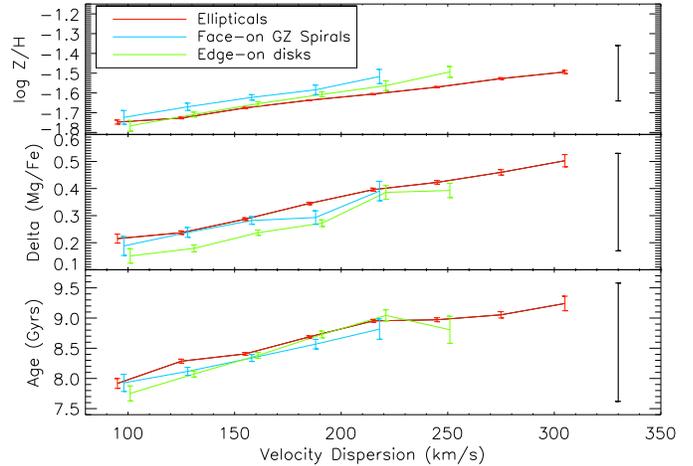,width=9cm} 
\caption{Stellar population properties as a function of the central $\sigma_v$.Lines and error bars are as in Fig.~\ref{fig:vsmass}.}
\label{fig:vssigma}
\end{center}
\end{figure}

The total stellar mass of a galaxy is a good tracer of its properties, but as we study the stellar populations in the {\it central regions} of the object, we perform a similar exercise in Fig.~\ref{fig:vssigma}, except that this time we show the metallicity, $\alpha-$enhancement and age of the stellar populations as a function of the measured velocity dispersion ($\sigma_v$) in the central region of the galaxy.
\begin{figure}
\begin{center}
\psfig{file=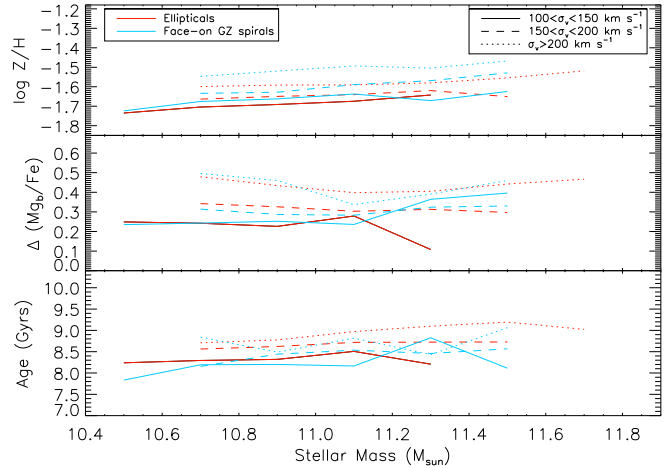,width=9cm} 
\caption{Stellar population properties as a function of total stellar mass, in bins of velocity dispersion. Trends with stellar mass are almost nonexistent, in comparison with \ref{fig:vsmass}, implying that the central velocity dispersion (a good proxy for the bulge's mass) and not {\it total} stellar mass is the observable with the strongest correlation with stellar population parameters.
}
\label{fig:sigbins}
\end{center}
\end{figure}

\begin{figure*}
\begin{center}
\psfig{file=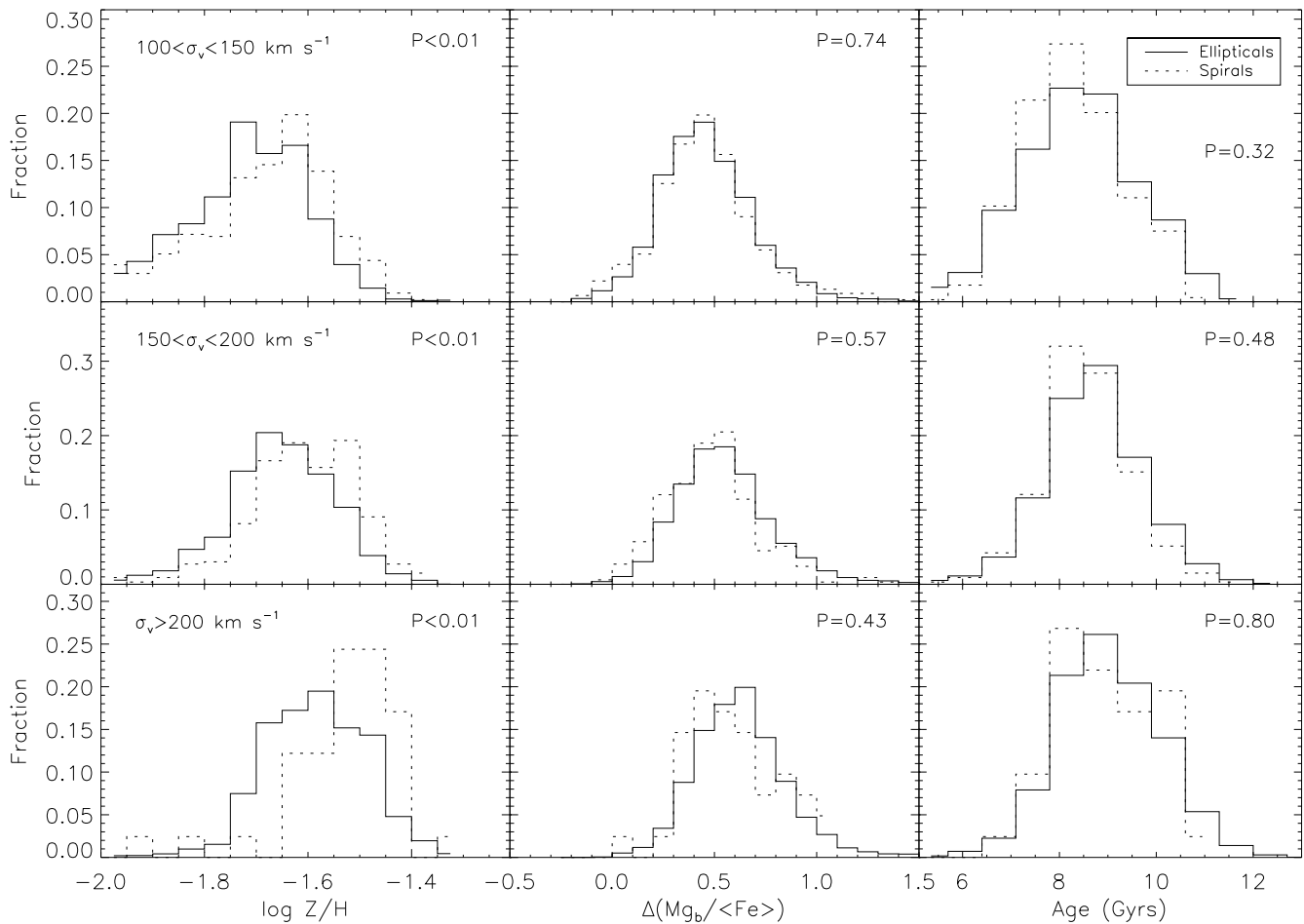,width=18cm} 
\caption{Distributions of metallicity, $\Delta$(Mgb/$\langle$Fe$\rangle$ and light-weighted age in different bins of central velocity dispersion for quiescent spiral (dotted line) and elliptical (solid line) galaxies. The probability of both samples to be drawn from the same distribution, obtained from a 2-sided Kolmogorov Smirnov test, is shown in each pannel. Bulges of quiescent spiral galaxies have higher metallicity than ellipticals with the same $\sigma_v$ at high confidence. $\alpha$-enhancement and age differences bewtween the two samples are only marginally detected when combining the three bins in $\sigma_v$. See text for more details.
}
\label{fig:hist}
\end{center}
\end{figure*}

\begin{figure*}
\begin{center}
\psfig{file=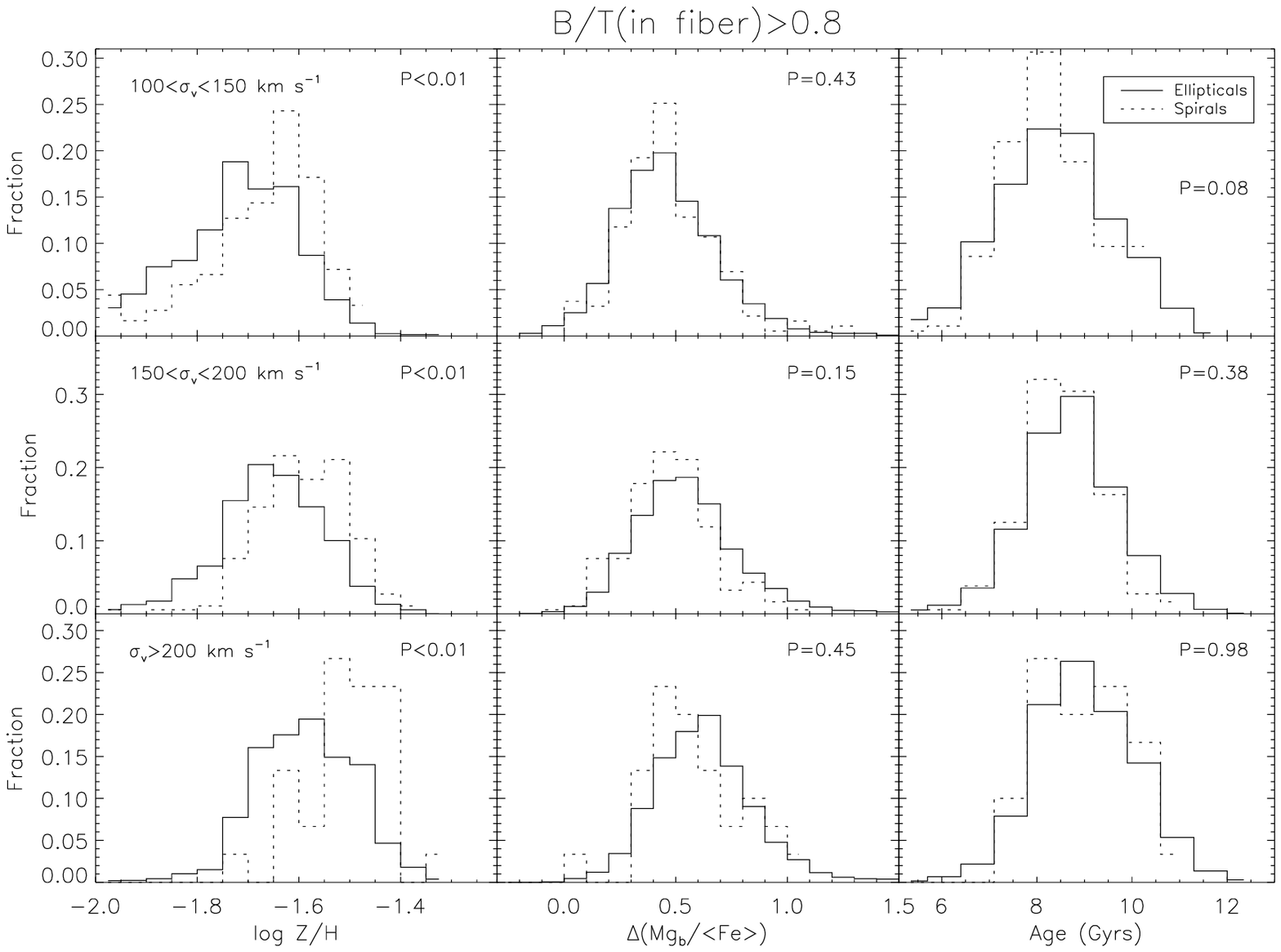,width=18cm} 
\caption{Same distributions as in Fig.~\ref{fig:hist} but using only galaxies with B/T$>0.8$ in the fiber --at least 80\% of the light contributing in the spectrum is produced by stars in the bulge. 
}
\label{fig:hist80}
\end{center}
\end{figure*}

 Velocity dispersion is not completely free from the influence of the mass distribution of the galaxy in the outer regions --i.e., outside the fiber coverage--, but has the advantages of being a direct measurement of the dynamical properties in the region of the galaxy under study and also less uncertainties related to modelling than total stellar mass. Nonetheless, while these quiescent spiral galaxies have prominent bulges, as interpreted from their high Sersic indexes, and are studied at radii smaller than their $R_e$, in some cases there will be some non-negligible contribution from the disk inside the fibre. 

In this case, the median  metallicity in the central regions of quiescent spirals is higher than that in early types. The measured $\Delta$(Mgb/$\langle$Fe$\rangle$) and light-weighted ages of spirals are consistent with the stellar populations of ellipticals at a given $\sigma_v$. The fact that at a given $\sigma_v$ all the three physical quantities re-scale in the same way with respect to ellipticals when comparing to Fig.~\ref{fig:vsmass}, indicates that the main source of the differences between those plots is the different B/T ratio in ellipticals and spirals. This makes us question which is the parameter which correlates the strongest with stellar populations; total stellar mass, or bulge mass (central velocity dispersion).
 
As shown in Fig.~\ref{fig:sigbins}, we repeat the exercise performed in Figs.~\ref{fig:vsmass} and \ref{fig:vssigma} but studying trends with stellar mass {\it in bins} of $\sigma_v$. In this case, there are only very weak (if at all) correlations between the age and $\alpha$-enhancement in our galaxies and their total stellar mass, implying that the velocity dispersion correlates the strongests with these parameters. However, it seems that there is a residual correlation between the metallicity and the stellar mass even when we factor out $\sigma_v$ --[Z/H] increases monotonically with stellar mass in all bins of $\sigma_v$ and for both morphological types.

We will from now on assume that central velocity dispersion is a good proxy for the mass of the bulge. The calculation of the mass of the bulge from a direct measurement of $\sigma_v$ can be, in principle, affected by the orbits of stars in the disk. However, \citet{cappellari06} use SAURON data to show that central $\sigma_V$ is proportional to $\sqrt{M}$ and weakly dependent on the orbital distributions. In addition we select only relatively face-on spiral galaxies, minimizing the impact of the disk's rotation on the $\sigma_v$ measurement.

The difference in total stellar mass, at a given bulge mass, must come mainly from the disk. The lack of a trend in age and $\alpha$ overabundance with total stellar mass, at fixed velocity dispersion, indicates that the stellar mass of the disk is not relevant in shaping these physical parameters. Instead, the residual trend in metallicity with total stellar mass at a given bulge mass implies that the extra mass of the disk might be relevant for the total metal content retained by the galaxy.

We now quantify the statistical significance of any possible difference between spirals and ellipticals by studying not only the median values of metallicity, $\Delta$(Mgb/$\langle$Fe$\rangle$) and age, but the whole distribution of these parameters in bins of $\sigma_v$. We show these distributions in Fig.~\ref{fig:hist}, where we have split our galaxy samples in three bins of central velocity dispersion. Using the 2-sided Kolmogorov-Smirnov (KS) test, we find that the bulges of quiescent spirals are more metal-rich than elliptical galaxies at fixed $\sigma_v$ with high confidence (probability of the two samples to be drawn from the same distribution $P\sim10^{-3}$). The difference in the median values is at the level of $\sim 0.07$ dex. While there are some differences in $\alpha-$enhancement and age (at least in some of the $\sigma_v$ bins), their statistical significance is not large, so physical interpretation is unnecessary, especially considering the typical uncertainties affecting the measurements.

\citet{anna08} identified the systematic uncertainties affecting the derivation of the stellar population parameters. The main contributions to the error budget in the case of metallicity are the lack of variation of [$\alpha$/Fe] in BC03 models, and to the choice of priors according to which the model library produces a galaxy's SFH\footnote{We refer the reader to \citet{anna08} for a full discussion on systematic uncertainties in the modelling.}. A combination of both factors can add an error of up to 0.046 dex in [Z/H], which is more than half of the difference we find between passive spirals and ellipticals. However, the fact that we see the same trend in all bins of $\sigma_v$ and the different shape of the distributions in Fig.~\ref{fig:hist} make us believe that there is a physical difference between the two galaxy samples. Moreover, we have shown that [$\alpha$/Fe] of spiral bulges and ellipticals are indistinguishable at fixed velocity dispersion, thus any bias introduced by $\alpha$/Fe would affect their total metallicity in a similar way and have negligible effect on our differential analysis.

So far, we have performed a differential analysis of the stellar populations in the {\it central regions} of spiral and elliptical galaxies, but we can not guarantee that all the light in the SDSS spectrograph fiber does indeed come from the bulge. In order to minimize this potential contamination from disk light we have chosen to work only with galaxies hosting {\it quiescent} disks, with stellar populations more similar to those in red and dead ellipticals than those in star forming disks. Furthermore it is possible to only select galxies for which the light in the fiber is dominated by the stars in the {\it bulge}. In order to do this we use the catalog by \citet{simard}, who have performed a bulge-disk decomposition for galaxies in the SDSS. We now choose only those galaxies for which at least $80\%$ of the light in the fiber is originated by bulge stars and repeat the analysis shown in Fig.~\ref{fig:hist}. It is worth noting that the total number of elliptical (spiral) galaxies is reduced by a 10\%(52\%).

Once again the metallicity seems to be the only discrepant quantity between the bulges of passive spirals and ellipticals at high significance ($P\sim10^{-3}$ in all three bins). The distribution of age and $\alpha$-enhancement in both galaxy samples are relatively similar. Visual inspection of the distributions, as well as the KS test probabilities, shows that the statistical evidence for a difference between the samples is inconclusive. High spatial resolution and 3D spectroscopy, rather than increasing sample size, would be more likely to test this conclusively.



Our findings point to a scenario in which the stellar population properties in the bulges of quiescent spiral and elliptical galaxies scale with the central velocity dispersion. We find the youngest, less metallic, less $\alpha$-enhanced objects to be those with the lower values of $\sigma_v$. This result is in good qualitative agreement with that of \citet{thomas06}, despite notable difference in sample selection and aperture definition. The main difference between the two works is the statistical significance reached because we use a sample $\sim 50$ times larger. Our larger sample size could also be driving our main discrepancy. While they find Z/H, $\alpha$-enhancement and age of spiral bulges to be equivalent to those in ellipticals at a given velocity dispersion, we find that the metallicity in bulges of quiescent spirals is higher than in elliptical galaxies.

However, it seems that the metallicity of those stellar populations {\it also} correlates with the total stellar mass of the galaxy. We know that there is a fundamental correlation between the mass of a galaxy and its metal content \citep[i.e.,][]{tremonti, anna05}, and that its origin is likely due to the higher capacity of more massive systems to retain metals in the presence of outflows.

 Taken together, our results imply that the formation epoch of galaxies and the duration of their star-forming period are linked to the mass of the bulge. The extent to which metals are retained within the galaxy, not being removed as a result of outflows, is determined by the total mass of the galaxy.

\citet{masters} studied the stellar populations of red spirals, finding them to be systematically older and with less recent SF activity than blue spiral galaxies. Combining their results with those presented here, imply a scenario in which SFHs of red spirals are more similar to those of ellipticals than those of star forming spirals.

We are very conservative in the selection of quiescent galaxies, so it is possible that a population of disk-like, low-level star formers do exist (and that is precisely what \citet{wolf09} and others report). However, a late quenching of the SF in disk galaxies could be reconciled with our results if, in spite of the relatively homogeneus color and lack of SF at $z \le 0.1$, there are radial gradients present in the stellar populations ages. In other words, if these galaxies grow inside-out \citep{barden} and the center of the objects, where the bulk of the stellar populations resides, was assembled at $z > 1$, the star formation per unit mass might be small enough to not to leave behind a strong imprint in the global colors of the object by $z \le 0.1$, and make it into our sample.

\citet{sauron} make use of SAURON survey data in order to study the stellar populations of 48 early-type galaxies. They find that the flattened component identified in fast-rotators does actually show an increase in the metallicity and a mildly depressed [$\alpha$/Fe] ratio with respect to the main body of the galaxy. Unfortunately their maps do not typically cover regions much larger than $R_e$, nor later morphological types than S0s. Future surveys, like the recently started Calar Alto Large Integral Field Area (CALIFA) survey will provide 3D spectroscopy over the full optical extent of a statistically significant sample of galaxies of all morphological types \citep{sanchez}, allowing to study the stellar populations of quiescent spirals at larger radii.

Assuming that similar $\alpha-$enhancements imply similar star formation timescales, it seems reasonable to believe that whatever the reason is after the shorter typical SF timescales  in elliptical galaxies, it is very likely that the stars in the bulges of quiescent spirals share a common formation mode with those in ellipticals. Furthermore, the fact that the light-weighted ages are similar at a given $\sigma_v$ implies that the epoch of the star formation shut-off must also be placed at the same epoch, and any subsequent episode of star formation must have happened at a lookback time high enough for any trace of young stellar populations to have dissapeared by today.

We would like to remind that while at stellar masses above $1-2 \times 10^{11}M_\odot$ the contribution of passive disks to the growth of the red sequence is very small \citep{arjen09}, probably because of a merger-dominated formation history at those masses \citep{arjen09,robaina10}, there are many passive disks contributing to such a growth since $z\sim 1$ \citep{bundy10, holden11} at lower masses. Therefore, it will be very important to understand when and how do spiral galaxies without noticeable star formation activity in the local Universe stopped forming stars. Future models of galaxy formation and evolution would have to be able to reproduce the small difference in metallicity

\section{Conclusions}

We have assembled a sample of galaxies with morphological classifications from Galaxy Zoo and photometry and light-profile fit parameters from the SDSS DR7, NYU-VAC. We perform a comparison between the stellar populations in the central regions of low inclination quiescent spiral galaxies and those in elliptical galaxies over the redshift range $0.04<z<0.1$ and for galaxies with stellar masses above $10^{10.4}M_\odot$. Specifically, we compare their r-band light-weighted ages, stellar metallicities and alpha-enhancement (as traced by $\Delta$(Mgb/$\langle$Fe$\rangle$)), derived as in G05 and G06.The main results of this analysis are:

\begin{itemize}

\item Central velocity dispersion is the observable with the strongest correlation with stellar population parameters. When we fix $\sigma_v$ we find no dependence of light-weighted age or $\alpha$-enhancement with total stellar mass for both elliptical and quiescent spiral galaxies. In the case of passive spirals, if we assume that central velocity dispersion is a good proxy for the bulge's mass, this implies that these parameters are independent of the disk's stellar mass. However, there is a residual correlation between [Z/H] and stellar mass even if we factor out central $\sigma_v$.

\item The metallicity of the stars in the bulges is higher in passive spirals than in ellipticals, at a given central velocity dispersion, by $\sim 0.07$ dex. On the other hand, median values and distributions of age and $\alpha$-enhancement are statistically compatible in both galaxy samples. This, together with the residual correlation found between metallicity and total stellar mass (bulge+disk), indicates a higher capacity of more massive systems to retain their metals during the process of star formation.


\item Our results are in good qualitative agreement with those of \citet{thomas06} in the case of age and $\alpha$-enhancement despite notable differences in sample selection and aperture definition. Our finding of a lower [Z/H] in quiescent ellipticals is likely a cause, given how small the difference is, of sample size.


\item As we are very strict with our criteria for selecting {\it passive} stellar populations, we are certainly not including in our sample all galaxies in transition from star-formers to passive-red. It is possible that our sample selection criteria are leaving out slightly less red disk galaxies with low levels of star formation and younger stellar populations, even if they are already on the red sequence. Models of galaxy formation have to account for a very large range in structural and morphological properties among galaxies with similar stellar ages and star formation timescales.

\end{itemize}

\section*{Acknowledgements}

The authors are thankful to the referee, whose comments and suggestions have greatly improved this work. A.~R.~R.~and L.~V.~acknowledge support from FP7-
IDEAS-Phys.LSS 240117. B.~H. acknowledges grant number FP7-PEOPLE-2007-4-3-IRG n 20218. The Dark Cosmology Centre is funded by the DNRF.

Funding for the Sloan Digital Sky Survey (SDSS) has been provided by the Alfred P. Sloan Foundation, the Participating Institutions, the National Aeronautics and Space Administration, the National Science Foundation, the U.S. Department of Energy, the Japanese Monbukagakusho, and the Max Planck Society. The SDSS Web site is http://www.sdss.org/.

The SDSS is managed by the Astrophysical Research Consortium (ARC) for the Participating Institutions. The Participating Institutions are The University of Chicago, Fermilab, the Institute for Advanced Study, the Japan Participation Group, The Johns Hopkins University, Los Alamos National Laboratory, the Max-Planck-Institute for Astronomy (MPIA), the Max-Planck-Institute for Astrophysics (MPA), New Mexico State University, University of Pittsburgh, Princeton University, the United States Naval Observatory, and the University of Washington.

This publication has been made possible by the participation of more than 160 000 volunteers in the Galaxy Zoo project. Their contributions are individually acknowledged at http://www.galaxyzoo.org/Volunteers.aspx

\end{document}